\documentstyle[12pt,epsf]{article}    % Simple LaTeX, use with DINA4 as follows
\newlength{\dinwidth}
\newlength{\dinmargin}
\def\lsim{\mathrel{\rlap{\lower4pt\hbox{\hskip1pt$\sim$}}
    \raise1pt\hbox{$<$}}}                % less than or approx. symbol
\def\gsim{\mathrel{\rlap{\lower4pt\hbox{\hskip1pt$\sim$}}
    \raise1pt\hbox{$>$}}}                % greater than or approx. symbol
\newcommand{\beq}  {\begin{equation}}
\newcommand{\eeq}  {\end{equation}}
\newcommand{\bmath}{\begin{eqnarray}}
\newcommand{\emath}{\end{eqnarray}}
\newcommand{\TS}   {\textstyle}

\begin{document}
\begin{titlepage}
\samepage{
\setcounter{page}{0}
\vspace{-1.in}
\rightline{Edinburgh 96/22}
\rightline{DFTT 58/96}
\rightline{hep-ph/9609515}
\vspace{.2in}
\begin{center}
{\bf  Measurement of the Polarized Structure Function $g_{1}^{p}(x,Q^{2})$ at
HERA \\}
\vspace{.5in}
R.~D.~Ball$^{a}$, A.~Deshpande$^{b,\,}\footnote{Royal Society University
Research Fellow.}$, S.~Forte$^{c}$,
V.~W.~Hughes$^{b,\,}\footnote{Supported by the Department of Energy.} $,
 J.~Lichtenstadt$^{d, e,\,}\footnote{Supported by the Israel
Science Foundation of the Israeli Academy of
Sciences.}$, G.~Ridolfi$^{f}$\\
\vspace{.2in}
 Department of Physics and Astronomy, Edinburgh EH9 3JZ, Scotland$^{(a)}$\\
 Department of Physics, Yale University, New Haven, CT 06511, USA$^{(b)}$\\
 INFN , Sezione di Torino, I-10125 Torino, Italy$^{(c)}$\\
 CERN, CH-1211 Geneva 23, Switzerland$^{(d)}$\\
 School of Physics and Astronomy, The Raymond and Beverly Sackler Faculty of
Exact\\ Sciences, Tel Aviv University, Tel Aviv 69978,  Israel$^{(e)}$
\\
 INFN, Sezione di Genova, I-16146, Genova, Italy$^{(f)}$\\

\end{center}
\vspace{.25in}

\begin{abstract}
We present estimates of possible data on spin-dependent asymmetries
in inclusive scattering of high energy polarized electrons by high
energy polarized protons at HERA, including statistical
errors, and discuss systematic uncertainties.
We show that these data would shed light on the small $x$ behaviour
of the polarized structure function $g_1$, and would reduce
substantially the uncertainty on the determination of the polarized
gluon distribution.
\end{abstract}
\vspace{.2in}
\begin{center}
  Presented at the 1996 HERA Workshop  \\
``Future Physics at HERA''\\
\vspace{.1in}
{\it To be published in the proceedings}
\end{center}
\vfill
%\leftline{CERN-TH/95-184}
\leftline{September 1996}}
\end{titlepage}
\setcounter{footnote}{0}
\newpage
\section{Introduction}
Nucleon structure, particularly as defined by its structure functions
determined from lepton-nucleon inclusive electromagnetic scattering,
is of fundamental importance and has provided crucial information for
the development of perturbative QCD.
The history of such experiments over the past forty years has shown that
important new information has been obtained when measurements were extended
to new kinematic regions. In the mid-1950's at Stanford,
Hofstadter~\cite{hof}
extended measurements of elastic electron-proton scattering to a higher
$Q^{2}$ range of 1 (GeV/c)$^{2}$ and first observed that the proton has
a finite size. In the late 1960's at SLAC Friedman, Kendall and
Taylor~\cite{FKT} extended measurements of inelastic inclusive electron
scattering to the deep inelastic region of $Q^{2}>1$~(GeV/c)$^{2}$ and
discovered the parton substructure of the proton.

\begin{figure}
\epsfxsize=10cm
\epsfysize=10cm
\hfil
\epsffile{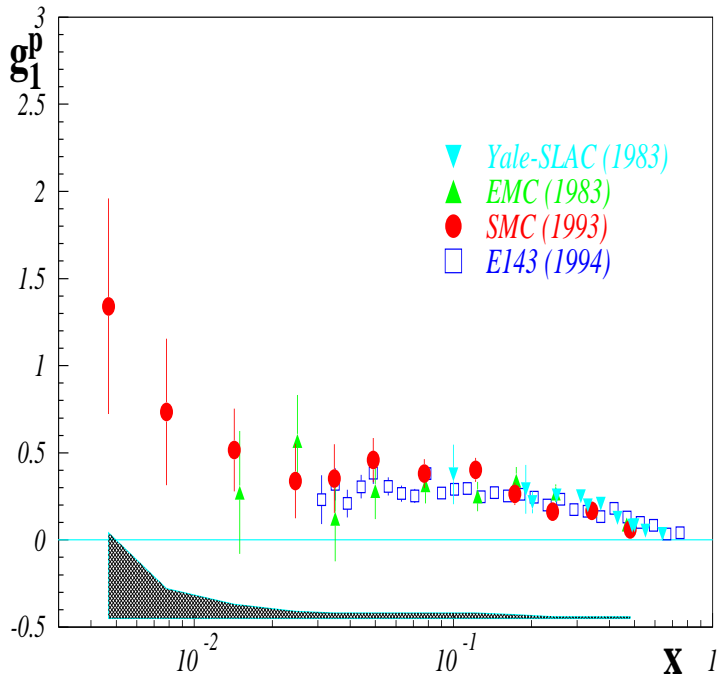}
\hfil
\caption{ {\it The current status of the measurement of the spin structure
function $g_{1}^{p}$. The statistical errors are shown with the data points,
while the size of the systematic errors for the SMC measurement is shown by the
shaded area.} }
\label{g1pstatus}
\end{figure}

The subfield of polarized lepton-proton scattering was initiated with the
Yale-SLAC E80 and E130 experiments~\cite{Y-SLAC} which
measured the spin-dependent structure
function of the proton in the mid-1980's. These experiments were then followed
up at CERN by the EM Collaboration which extended the kinematic range of the
original measurements at SLAC to lower the $x$ range from $x=0.1$ to $0.01$
\cite{EMC}.
These data allowed a determination of the singlet component of the first
moment of $g_1$, which in the naive parton model is the fraction
of the proton spin carried by quarks: this was found to be
compatible with zero, thereby violating the parton model Ellis-Jaffe
sum rule at the three standard deviation level.
%This surprising result has stimulated a large amount of experimental
This surprising result has stimulated a large amount of
experimental
and theoretical work on polarized  structure functions~\cite{spinrev}.
New measurements of the spin dependent structure functions of the proton
and the deuteron were made
by the SM collaboration~\cite{SMC_papers} at CERN and by
E143~\cite{E143_papers}
at SLAC which extended the kinematic range to lower $x$ and also reduced
the statistical and systematic uncertainties significantly.
Figure \ref{g1pstatus} shows all published measurements of the structure
function
$g_{1}^{p}$.
These measurements, along with those made for the deuteron,
allowed a better determination of the first moment of the spin structure
functions
and a verification of the Bjorken sum rule. Furthermore, they made
a next-to leading order QCD analysis of the $x$ and $Q^2$ dependence of
$g_1$ possible, thereby
allowing a determination of the first moment of the polarized gluon
distribution~\cite{BFRa}.

In view of all this, it is important to consider in detail what
measurements and accuracies may be obtained at HERA. The
primary goal and motivation of HERA is to extend the
kinematic range of electron-proton scattering. Thus far HERA research and its
discoveries were the result of the new and vastly extended kinematic range in
$x$ and $Q^{2}$ , both by a factor of about 100, provided by the collider
\cite{deroeck,rdbdr}. At present the $800$~GeV
proton beam at HERA is unpolarized (although the
$25$~GeV electron/positron beam has a natural polarization).
If the proton beam were to be polarized, it would be possible
to measure polarization asymmetries and thus explore spin
dependent structure functions along with the spin-independent structure
functions in the HERA kinematic range. In this paper we discuss what
might be achieved from such a programme. We show that
the data HERA could provide are clearly unique and cannot be
obtained by present day experiments.

In section 2 we estimate the data that could be obtained, including the
statistical errors, from inclusive polarized e-p scattering with HERA in
the collider mode. We also discuss systematic errors and argue that they can be
controlled within adequate limits. In section 3 we briefly review the
relation of the structure function $g_1$ to polarized parton
distributions and the perturbative evolution equations satisfied by the latter,
and summarize the status of polarized parton distributions
extracted from presently available fixed target data.
In section 4 we show that new data from polarized colliding beam
experiments at
HERA would shed light on the small $x$ behaviour of $g_1$ and would
substantially improve the determination of the polarized gluon distribution.
%\newpage

\section{Measurement of $g_{1}^{p}(x,Q^{2})$ at HERA }

\subsection{Kinematic range and statistical errors for HERA data}

\begin{figure}
\epsfxsize=10cm
\epsfysize=10cm
\hfil
\epsffile[20 20 540 540]{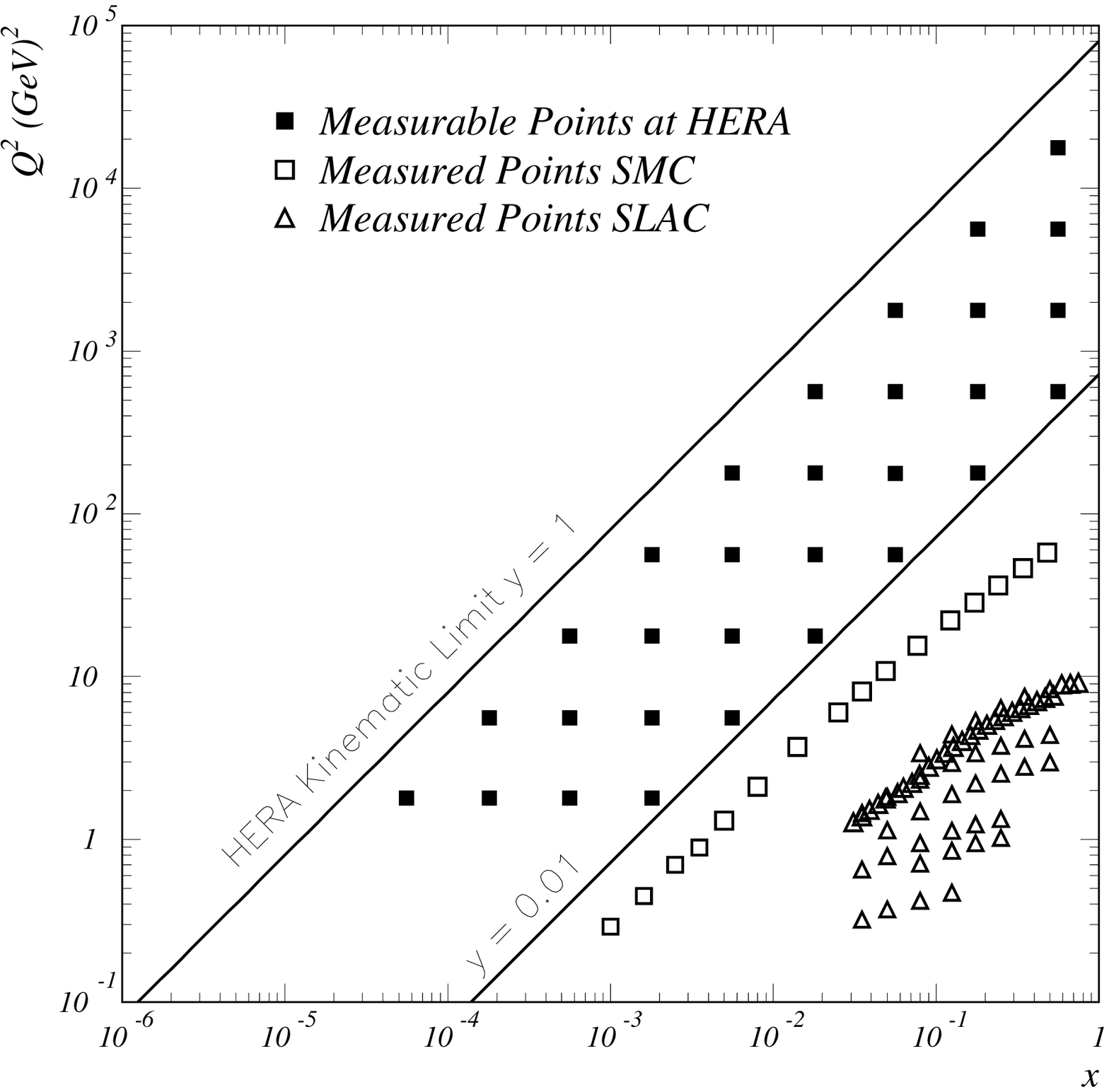}
\hfil
\caption{ {\it  Measurable $x-Q^{2}$ region at HERA shown with the presently
explored regions by SMC(CERN) and E143(SLAC) experiments, and the kinematic
limit of measurability at HERA.}}
\label{xq2_hera}
\end{figure}

Presently all measurements of polarized structure functions are made
using fixed target deep inelastic lepton-nucleon scattering.
Figure \ref{xq2_hera} shows the kinematic ranges in which polarized
lepton-proton data have been obtained at SLAC and CERN as well as the possible
measurements in the kinematic domain available at HERA with 800 and 25 GeV
proton and electron beams, respectively. Note the large extension in $x-Q^{2}$
range compared to the present data. Table 1 lists the $x-Q^{2}$ values at which
measurements can be made at HERA, the number of events that would be measured
and the associated statistical errors in the measured asymmetry $\delta A_{m}$
assuming an integrated luminosity of $L=1000$~pb$^{-1}$ and electron and proton
polarizations of $0.7$ each\footnote{ Accelerator parameters that
were suggested by R.~Klanner and F.~Willeke for this workshop.}.
Kinematic cuts~\cite{deroeck} on $y$ and the scattered electron angle
$\theta_{e}'$ used for data analysis by the H1 and ZEUS collaborations
were applied for the evaluation of the counting rates
in each $x-Q^{2}$ bin. Standard deep inelastic scattering formulae were used
to calculate of the kinematics and asymmetries. They are given in the
Appendix.
\begin{table}
\hfil
\begin{tabular}{|c|c|c|c|c|c|}\hline\hline
$x$ & $Q^{2}$ GeV$^{2}$ & $y$ & $D$ & $N_{total}$  & $\delta A_{m}$ \\
\hline\hline
$5.6\times 10^{-5}$ & $1.8$ & $0.40$ & $0.47$ & $5.0\times 10^{6}$ & $4.5\times
10^{-4}$ \\
\hline
$1.8\times 10^{-4}$ & $1.8$ & $0.13$ & $0.13$ & $8.0\times 10^{6}$ & $3.5\times
10^{-4}$ \\
                    & $5.6$ & $0.40$ & $0.46$ & $3.5\times 10^{6}$ & $5.3\times
10^{-4}$ \\
\hline
$5.6\times 10^{-4}$ & $1.8$ & $0.04$ & $0.04$ &$1.2\times 10^{7}$ & $2.9\times
10^{-4}$ \\
                    & $5.6$ & $0.13$ & $0.13$ &$6.5\times 10^{6}$ & $3.9\times
10^{-4}$ \\
                    & $1.8\times 10^{1}$ & $0.40$ & $0.47$ & $2.0\times 10^{6}$
& $7.1\times 10^{-4}$ \\
\hline
$1.8\times 10^{-3}$ & $1.8$ & $0.01$ & $0.01$ & $1.6\times 10^{7}$ & $2.5\times
10^{-4}$ \\
                    & $5.6$ & $0.04$ & $0.04$ & $9.0\times 10^{6}$ & $3.3\times
10^{-4}$ \\
                    & $1.8\times 10^{1}$ & $0.13$ & $0.13$ & $3.4\times 10^{6}$
& $5.4\times 10^{-4}$ \\
                    & $5.6\times 10^{1}$ & $0.40$ & $0.47$ & $1.1\times 10^{6}$
& $9.5\times 10^{-4}$ \\
\hline
$5.6\times 10^{-3}$ & $5.6$ & $0.01$ & $0.01$ & $1.2\times 10^{7}$ & $2.8\times
10^{-4}$ \\
                    & $1.8\times 10^{1}$ & $0.04$ & $0.04$ & $5.5\times 10^{6}$
& $4.3\times 10^{-4}$ \\
                    & $5.6\times 10^{1}$ & $0.13$ & $0.13$ & $1.7\times 10^{6}$
& $7.7\times 10^{-4}$ \\
                    & $1.8\times 10^{2}$ & $0.40$ & $0.47$ & $5.5\times 10^{5}$
& $1.3\times 10^{-3}$ \\
\hline
$1.8\times 10^{-2}$ & $1.8\times 10^{1}$ & $0.01$ & $0.01$ & $6.5\times 10^{6}$
& $3.9\times 10^{-4}$ \\
                    & $5.6\times 10^{1}$ & $0.04$ & $0.04$ & $2.6\times 10^{6}$
& $6.2\times 10^{-4}$ \\
                    & $1.8\times 10^{2}$ & $0.12$ & $0.13$ & $8.0\times 10^{5}$
& $1.1\times 10^{-3}$ \\
                    & $5.6\times 10^{2}$ & $0.40$ & $0.47$ & $2.3\times 10^{5}$
& $2.1\times 10^{-3}$ \\
\hline
$5.6\times 10^{-2}$ & $5.6\times 10^{1}$ & $0.01$ & $0.01$ & $2.2\times 10^{6}$
& $6.7\times 10^{-4}$ \\
                    & $1.8\times 10^{2}$ & $0.04$ & $0.04$ & $8.0\times 10^{5}$
& $1.1\times 10^{-3}$ \\
                    & $5.6\times 10^{2}$ & $0.12$ & $0.13$ & $2.6\times 10^{5}$
& $2.0\times 10^{-3}$ \\
                    & $1.8\times 10^{3}$ & $0.40$ & $0.47$ & $6.5\times 10^{4}$
& $3.9\times 10^{-3}$ \\
\hline
$1.8\times 10^{-1}$ & $1.8\times 10^{2}$ & $0.01$ & $0.01$ & $5.9\times 10^{5}$
& $1.3\times 10^{-3}$\\
                    & $5.6\times 10^{2}$ & $0.04$ & $0.04$ & $1.9\times 10^{5}$
& $2.3\times 10^{-3}$\\
                    & $1.8\times 10^{3}$ & $0.13$ & $0.13$ & $5.4\times 10^{4}$
& $4.3\times 10^{-3}$\\
                    & $5.6\times 10^{3}$ & $0.40$ & $0.47$ & $1.3\times 10^{4}$
& $8.6\times 10^{-3}$\\
\hline
$5.6\times 10^{-1}$ & $5.6\times 10^{2}$ & $0.01$ & $0.01$ & $2.9\times 10^{4}$
& $5.9\times 10^{-3}$\\
                    & $1.8\times 10^{3}$ & $0.04$ & $0.04$ & $8.3\times 10^{3}$
& $1.1\times 10^{-2}$\\
                    & $5.6\times 10^{3}$ & $0.13$ & $0.13$ & $1.9\times 10^{3}$
& $2.3\times 10^{-2}$\\
                    & $1.8\times 10^{4}$ & $0.40$ & $0.47$ & $4.2\times 10^{2}$
& $4.8\times 10^{-2}$\\
\hline\hline
\end{tabular}
\hfil
\label{tab-asym}
\caption{ {\it   The kinematic variables $x,Q^{2},y,$ and $D$, along with the
number of events expected $N_{total}$, and the statistical uncertainty in
the measured asymmetry $\delta A_{m}$ in the kinematical region assuming
an integrated luminosity {\cal L} $=1000~pb^{-1}$ and proton and electron
beam polarizations $P_{p}=P_{e}=0.7$. }}
\end{table}
\subsection{Systematic errors for HERA data}
The systematic errors associated with spin dependent asymmetry measurements
are of two types: 1) normalization errors and 2) false asymmetries.

The normalization errors include principally uncertainties in the electron
polarization $P_{e}$ and in the proton polarization $P_{p}$ . These lead to a
change in the magnitude of the measured asymmetry but in practice by an amount
which is small compared to the statistical error, and hence they are important
primarily when evaluating the first moment of $g_{1}^{p}(x)$.

Recently the HERMES collaboration has reported
a measurement of $P_{e}$ by Compton scattering, which yielded
a relative accuracy of $5.5\%$
\cite{Hermese_DIS96}. It is expected that an
improved accuracy will be achieved eventually.
%%%%%%%%%%%%%%%%%%%%%%%%%%%%%%%%%%%%%%%%%%%%%%%%%
%  Patch re-written by Vernon 14 May 1996 Faxed by Pat.
%%%%%%%%%%%%%%%%%%%%%%%%%%%%%%%%%%%%%%%%%%%%%%%%%
%M\"{o}ller scattering \cite{Hermese_DIS96}.
%
%Measurement of the polarization of the high energy proton beam can be done
%using Coulomb nuclear interference (CNI) in p-p elastic scattering for
%an absolute determination\cite{CNI} and using inclusive pion production for
%%the
%relative determination\cite{Primak}.
%Thus far the only measurement of the polarization of a
%high energy proton beam has been done at Fermilab (E704) for $E_{p}~=200~GeV$.
%For their polarized proton beam from $\Lambda$-decay they measured
%$P_{p} = ??$ \cite{E704}.
%
%The method of measuring $P_{p}$ at $1~TeV$ has been discussed in some detail
%in a design report entitled ``Acceleration of Polarized Protons to $120~GeV$
%and $1~TeV$ at Fermilab''~\cite{PpFermi}
%and also in the
%BNL proposal for RHIC SPIN~\cite{RHICSPIN}. The conclusion is that
%measurement to a relative
%accuracy of about $5\%$ should be possible.

Absolute measurement of the polarization of the high energy (800~GeV) proton
beam presents a new challenge and is presently under investigation.
Several methods are being considered:
\begin{itemize}
\begin{enumerate}
\item
$p-p$ elastic scattering in the Coulomb-nuclear interference region\cite{CNI},
\item $p-\vec {\rm p}$ scattering using a stationary polarized proton
jet target and comparing the asymmetries $A_{\rm beam} = A_{\rm jet}$,
where $A_{\rm beam}$ is the
asymmetry in the scattering of the polarized beam
%with
%polarization $P_{\rm beam}$
from an $unpolarized$ jet
and $A_{\rm jet}$ is the asymmetry measured with
an $unpolarized$ beam and a polarized jet target
with a known polarization $P_{\rm jet}$. Adjusting the
jet polarization to obtain the same asymmetry, one obtains
the beam polarization $P_{\rm beam} = P_{\rm jet}$~\cite{Wolfenstein}.
%\item p-p scattering using a stationary polarized proton jet target and using
%the relation that A=P in which A is the
%asymmetry in the scattering of the polarized beam with
%polarization P from an unpolarized jet and P is the polarization of a jet
%which produces the same asymmetry from scattering of an unpolarized beam
%\cite{Wolfenstein}.
\item polarized $e-p$ scattering in which a low energy polarized electron beam
collides with the polarized 800~GeV proton beam, with kinematics corresponding
to the polarized $e-p$ elastic scattering measurements at SLAC \cite{alg76}.
The theoretical values for
the asymmetry in elastic scattering are given in terms of the
polarizations and the measured electric
and magnetic form factors of the proton. Hence the measured asymmetry at HERA
using a known polarized electron beam would determine $P_{p}$.
\end{enumerate}
\end{itemize}

We comment briefly on these possible methods. For method 1 the nuclear matrix
element may not be known well enough to provide a significant absolute
determination
of $P_{p}$. Method 2 appears quite attractive because only the invariance
principle is needed from theory, and it may be possible to use the stationary
polarized target of the HERMES experiment for this measurement.
Method 3 is theoretically
sound and simple, but may not be attractive experimentally.
These methods have still to be fully developed and tested experimentally.
%Much further serious
%design effort is required for all these methods.
It is reasonable to expect
that a 5\% accuracy will be achieved.

Inclusive $\pi^{+}$ or $\pi^{-}$ production in collisions of $800$~GeV protons
with a fixed proton target appears to be a simple method for a relative
measurement of the polarization \cite{adams91}.

Methods to measure $P_{p}$ are presently considered and studied also for
polarized protons at RHIC.
%Considerable thought and is currently devoted to these methods to
%measure $P_{p}$.
The method of measuring $P_{p}$ at $1$~TeV has been discussed in
some detail in a design report entitled ``Acceleration of Polarized Protons
to $120$~GeV and $1$~TeV at Fermilab''~\cite{PpFermi} as well as in the
BNL proposal for RHIC SPIN~\cite{RHICSPIN}.

The other type of systematic error comes from false asymmetries.
These arise from variations in
counting rate due to time variations in detector efficiencies, beam
intensities,
or crossing angles between the conditions of spins parallel and antiparallel.
In view of the expected small values of the true asymmetries expected in the
HERA kinematic region,
(Table 1),
the false asymmetries must be controlled at the level of
$10^{-4}$.
We note that medium and high energy experiments which have measured
parity violation
have controlled false asymmetries to less than $10^{-5}$ to $10^{-8}$
\cite{parvioExp}.

False asymmetries can be avoided by frequent reversals of spin orientations.
In the HERA collider the electron(positron) spin reversal is difficult
and time consuming (several hours) and would lead to changes in
beam intensity and beam
emittance. Hence asymmetry data must be obtained with a fixed helicity
for the electron beam by varying the proton helicity. Both the electron and
proton rings are filled with about 200 particle bunches. The proton ring can be
filled with bunches of protons with individually predetermined
polarizations.
Parallel and antiparallel spin
data would be obtained at successive beam crossings occurring at time
intervals of about 100 ns.
Such  alternations eliminate many errors in an asymmetry measurement. However,
differences in the intensities and the crossing angles at successive proton
bunches could still lead to false asymmetries and will have to be minimized.

The most important approach to avoid false asymmetries will be to use a spin
rotator to reverse the helicity of all proton bunches at adequately
frequent intervals (perhaps once per 8 hours).
The design of such rotators has been studied and appears
practical without appreciable change in the orbits, or in the magnitude
of proton polarization~\cite{Roser}. Studies at the IUCF at $370$~MeV have
confirmed
these conclusions~\cite{Phelps}. With such approach the false asymmetries could
be
controlled to less than $10^{-4}$.

%\newpage

\section{Current status of $g_1(x,Q^2)$}

\subsection{Perturbative evolution of $g_1(x,Q^2)$}
\label{nlo}
The definition and properties of the polarized structure function
$g_1$ in perturbative QCD closely parallel that of its unpolarized
counterpart $F_2$ (see ref.~\cite{alta82} for a review).
The structure function
$g_1$ is related to the polarized quark and gluon distributions through
\bmath
   g_1(x,t) & = & {\TS\frac{1}{2}} \langle e^2\rangle
        \int_x^1 \frac{{\rm d}y}{y} \Bigl [
        C_q^{\rm S}({\TS\frac{x}{y}},\alpha_s(t)) \Delta\Sigma(y,t)
      \nonumber \\
      & + & 2 n_f C_g({\TS\frac{x}{y}},\alpha_s(t)) \Delta g(y,t)
    + C_q^{\rm NS}({\TS\frac{x}{y}},\alpha_s(t)) \Delta q^{\rm NS}(y,t) \Bigr],
\label{qcd:g1}
\emath
where $\langle e^2\rangle = n_f^{-1}\sum_{k=1}^{n_f} e_k^2$,
$t = \ln(Q^2/\Lambda^2)$,
$\Delta\Sigma$ and $\Delta q^{\rm NS}$ are the singlet and non-singlet
polarized quark distributions
$$
\Delta\Sigma(x,t) =  \sum_{i=1}^{n_f} \Delta q_i(x,t),
\qquad
\Delta q^{\rm NS}(x,t)  =  \sum_{i=1}^{n_f}
                ( e_i^2/\langle e^2\rangle -1)\Delta q_i(x,t),
\label{qcd:qsns}
$$
and $C_q^{S,NS}(\alpha_s(Q^2))$ and $C_g(\alpha_s(Q^2))$ are
the quark and gluon coefficient functions.

The $x$ and $Q^2$ dependence of the polarized quark and gluon distributions
is given by Altarelli-Parisi equations~\cite{AP}:
\bmath
{{\rm d} \over {\rm d} t} \Delta\Sigma(x,t) & = & \frac{\alpha_s(t)}{2\pi}
     \int_x^1 \frac{{\rm d}y}{y} \left [
     P_{qq}^{\rm S}({\TS\frac{x}{y}},\alpha_s(t)) \Delta\Sigma(y,t)
     + 2 n_f P_{qg}({\TS\frac{x}{y}},\alpha_s(t)) \Delta g(y,t) \right ],
\label{qcd:apsi}
\\
{{\rm d} \over {\rm d} t} \Delta g(x,t)  & = & \frac{\alpha_s(t)}{2\pi}
     \int_x^1 \frac{{\rm d}y}{y} \left [
     P_{gq}({\TS\frac{x}{y}},\alpha_s(t)) \Delta\Sigma(y,t)
     + P_{gg}({\TS\frac{x}{y}},\alpha_s(t)) \Delta g(y,t) \right ],
\label{qcd:apg}
\emath
\beq
   {{\rm d} \over {\rm d} t} \Delta q^{\rm NS}(x,t) =
      \frac{\alpha_s(t)}{2\pi} \int_x^1 \frac{{\rm d}y }{y}\,
      P_{qq}^{\rm NS}({\TS\frac{x}{y}},\alpha_s(t)) \Delta q^{\rm NS}(y,t),
\label{qcd:apns}
\eeq
where  $P_{ij}$ are polarized splitting functions.

The full set of  coefficient functions \cite{cf} and splitting functions
\cite{Pij} has been computed up to next-to-leading order in $\alpha_{s}$.
As in any perturbative calculation, at next-to-leading order
splitting functions, coefficient functions
and parton distributions depend on renormalization
and factorization scheme, while of course physical observables,
such as $g_1$ itself, remain scheme-independent up to terms of
order $\alpha_s^2$. The scheme choice is arbitrary, and in
particular parton distributions in
different factorization schemes are related to each other by well-defined
linear transformations.

The factorization scheme dependence
is particularly subtle in the polarized case
because of the extra ambiguity related to the definition of the $\gamma_5$
matrix, i.e. to the way chiral symmetry is broken by the regularization
procedure. This is reflected in an ambiguity in the
size of the first moment of the gluon coefficient function $C_g$
(which starts at order $\alpha_s$).
Two widely adopted choices~\cite{BFRa,SG,GRSV},
both compatible with the choice of $\overline{\rm MS}$
renormalization and factorization, correspond to either
requiring the first moment of the gluon coefficient function to be $C_g^1=0$,
or imposing that the first moment of the polarized quark distribution
be scale independent, which implies $C_g^1=-{\alpha_s\over 4\pi}$.
The first moment $\Delta g^1$ of the gluon distribution can be chosen to be
the same in the two schemes, whereas the first moments of the quark
distribution in the two schemes differ by an
amount proportional to $\alpha_s \Delta g^1$. Because the
evolution equations \ref{qcd:apsi}-\ref{qcd:apns}
imply that at leading order the first moment of the
polarized gluon scales as ${1\over \alpha_s}$
this scheme dependence persists asymptotically and is potentially large
if the first moment of the gluon distribution is large~\cite{anom}.

\subsection{Current status of polarized parton distributions.}

Parton distributions can be extracted from experimental structure function
data by parametrizing them at a starting value of $Q^2$, evolving this initial
condition up to any desired value of $x$ and $Q^2$ using
Eqs.~\ref{qcd:apsi}-\ref{qcd:apns},
determining $g_1$ there by means of
Eq.~\ref{qcd:g1}, and determining the initial parametrization which
gives the best fit of $g_1(x,Q^2)$ to the data~\cite{BFRa,SG,GRSV}.
Here we follow the procedure used in
refs.~\cite{BFRa,BFR}: we give the initial conditions at
$Q^2=1$~GeV$^2$ in the form
\begin{equation}
\label{part_dist}
\Delta f(x,Q^{2}) =
N(\alpha_f,\beta_f,a_f)~\eta_f~x^{\alpha_f}(1-x)^{\beta_f}(1+a_f~x),
\end{equation}
where $N(\alpha,\beta,a)$ is fixed by the normalization condition,
$N(\alpha,\beta,a)\int_{0}^{1}dx x^{\alpha}(1-x)^{\beta}(1+ax)=1$, and $\Delta
f$ denotes $\Delta \Sigma$, $\Delta q_{NS}$, or $\Delta g$.
With this normalization the parameters
$\eta_{g},\eta_{NS}$, and $\eta_{S}$ are respectively
the first moments of the gluon, the
non-singlet quark and the singlet quark distributions at
the starting scale. Evolution is performed within
the AB factorization scheme
(which has $C_g^1=-{\alpha_s\over 4\pi}$).
Further details of the fits and analysis are given in ref.~\cite{BFRa,BFR}.

\begin{figure}
\epsfxsize=14cm
\epsfysize=14cm
\hfil
\epsffile[0 280 510 510]{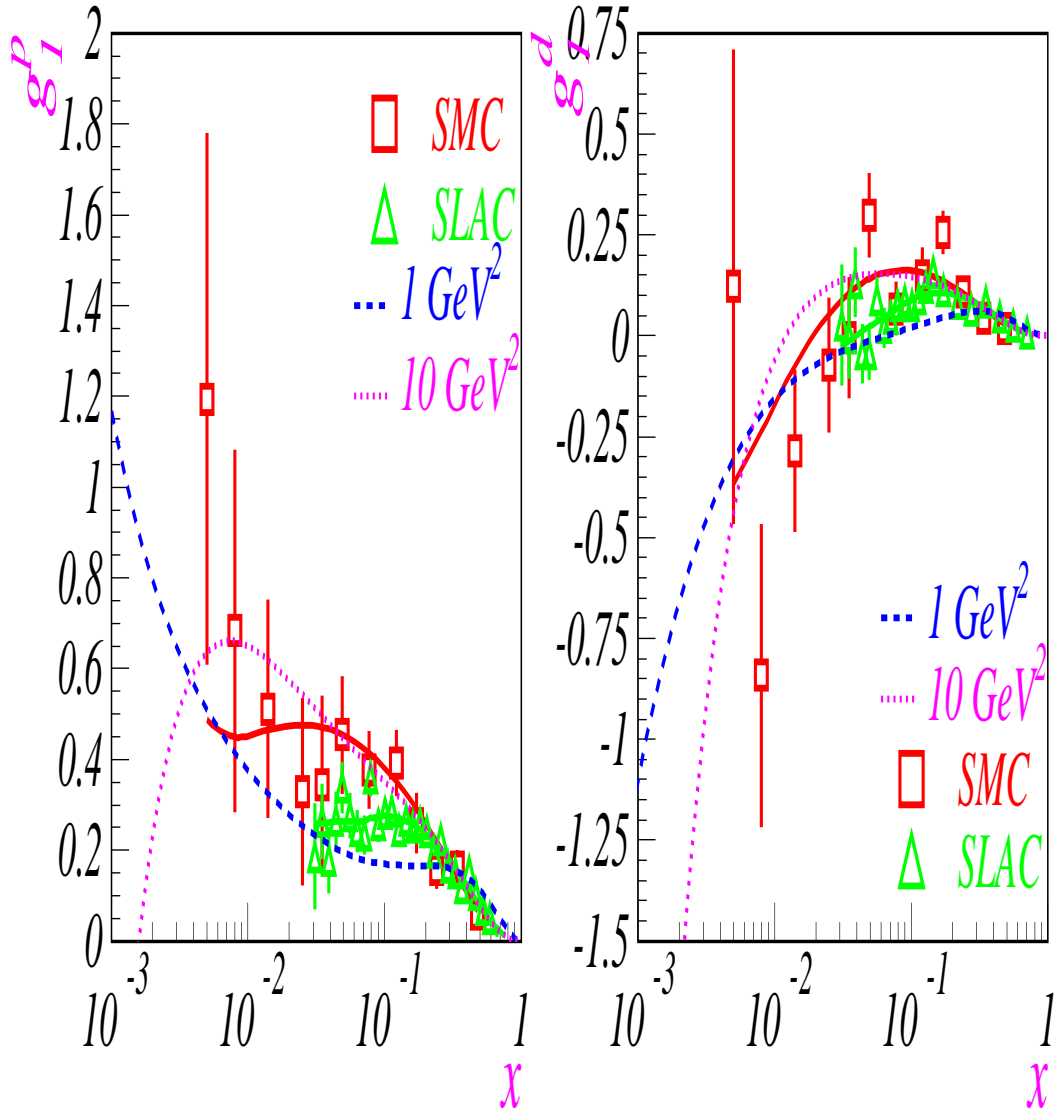}
\hfil
\caption { {\it The NLO fit to proton $g_{1}^{p}$ and
deuteron $g_1^{\rm d}$ data. The solid lines are fits to data at the
measured $Q^{2}$ values, and
the dashed and the dotted lines are fits evolved to $Q^{2}=$ 1
and 10 GeV$^{
2}$ respectively.} }
\label{figbfrpd}
%\caption{ {\it   BFR NLO FITS to proton and deuteron structure functions
%%measurements}
\end{figure}

The results of the NLO fit of Ref.~\cite{BFRa} (based on the data
of Ref.s~\cite{SMC_papers, E143_papers})
are shown in Fig.~\ref{figbfrpd}.
Interestingly, the data require the first moment of the gluon distribution
to differ significantly from zero:
$\eta_g = 1.52\pm0.74$.  This result follows mostly from
the observed scaling violations in
the intermediate and small $x$ region. However,
the statistical uncertainty
on the size of the polarized gluon distribution is still rather large.
Moreover,
existing  data only partially constrain the small $x$  behaviour of the various
parton distributions (values of $\alpha_{S},\alpha_{g}$, and $\alpha_{NS}$),
and
do not allow a precise determination of their asymptotic form for small $x$.
Such information, besides its intrinsic theoretical interest, is required in
order to obtain a precise determination of the moments of
$g_1$~\cite{BFRa,BFR}. New data with an extended kinematic
coverage in $x$ and $Q^2$ provided by HERA could reduce these uncertainties.

Since the publication of ref.~\cite{BFRa}, new data on the $Q^{2}$
dependence of $g_1$ have been published by the E143
collaboration\cite{newE143}.
The inclusion
of these data in the NLO fits\footnote{We have
excluded data from ref.~\cite{newE143} corresponding to the beam energy
of 29.1~GeV, which   duplicate previously published data
\cite{E143_papers}, as well as data with $Q^2< 0.95$~GeV$^2$.},
results in values of the
fitted parameters (shown in column 2 of Table 2)
whose  central values
are consistent with
the previously published ones~\cite{BFRa}, while
errors are reduced of up to 20\%.
The value of the first moment of the gluon distribution at
$Q^2=1$~GeV$^2$ in particular becomes $\eta_g=1.30\pm 0.56$.
In the next two years
both SMC~(CERN) and E143~(SLAC) plan to present new data on proton and deuteron
spin structure functions and using these data as well a more accurate
determination of the parameters in the fit can be expected. It can thus be
anticipated that eventually the dominant uncertainty in the determination of
the first moments of $g_1$ and the polarized gluon distribution will be due to
lack of experimental information in the HERA region.
%\newpage

\section{The impact of polarized HERA data}

\subsection{The small $x$ behaviour of $g_1$}

As shown in Fig.~\ref{xq2_hera}, measurements of $g_1$ at HERA will
extend the $x$ region
down to $x=5.6\times 10^{-5}$. Knowledge of the small $x$ behaviour of $g_1$
is obviously necessary in order to compute moments of $g_1$, and indeed
in the determination of the singlet component of the first moment of $g_1$ the
uncertainty related to the lack of knowledge of this behaviour is already
comparable to the statistical uncertainty ~\cite{BFRa,BFR}.

The extrapolation of $g_1$ from the measured region down to $x=0$
is traditionally done by assuming Regge behaviour of the structure function,
which implies~\cite{heim} $g_1\sim x^\alpha$ as $x\to 0$ with $0\le \alpha
\le 0.5$, i.e. a valence-like behaviour of $g_1$.
This behaviour seems  to disagree with the
data (see Table~2 and Fig.~\ref{figbfrpd})
which suggest instead that the magnitude of both the singlet and nonsinglet
components of $g_1$ increase at small $x$.

In fact, a valence-like behaviour
of $g_1$ is incompatible with perturbative QCD, which at leading order
 predicts instead
that $g_1$ should rise in magnitude at least as $g_1(x,Q^2)\sim
{1\over\sqrt{\sigma}}e^{2\gamma\sigma}$,
where $\sigma\equiv\sqrt{\xi\zeta}$, $\rho\equiv\sqrt{\xi/\zeta}$,
$\xi\equiv\ln{x_0\over x}$,
$\zeta\equiv\ln{\alpha_s(Q_0^2)\over\alpha_s(Q^2)}$,
and $Q_0$ and $x_0$ are  reference values of $x$ and $Q^2$.
This rise is present both in the singlet and nonsinglet components of $g_1$,
but with different slopes $\gamma$,  calculable in perturbative QCD.
The sign of this rise depends on the specific form of the quark and
gluon distributions, but for most reasonable forms of $\Delta q$ and
$\Delta g$, and in particular if $\Delta g$ at moderately
small $x$ is positive definite, then $g_1$ will be negative. The onset
of this behaviour as $Q^2$ is raised is clearly shown in Fig.~\ref{figbfrpd}.
Higher order corrections lead to an even stronger drop:
at $k$-th perturbative order  the rise of $-g_1$ is enhanced by a factor of
$\alpha_s^k\rho^{2k+1}$. It has been suggested~\cite{doublelog}
that these terms to
all orders in $\alpha_s$ may
exponentiate, thus leading to a rise of $g_1$ as a power of $x$; the
sign of this rise is still predicted to be negative.

A non-Regge behaviour of the unpolarized structure function $F_2$
has been observed and accurately measured at HERA, in spectacular agreement
with the perturbative QCD prediction~\cite{deroeck,rdbdr}. The behaviour
correspondingly predicted in the  polarized case
is even more interesting due to the fact that higher order corrections
are stronger, and also the polarized singlet
and nonsinglet quark and gluon distributions all display qualitatively
the same behaviour, whereas in the unpolarized case only the gluon
dominates at small $x$.

\begin{figure}
\epsfxsize=9cm
\epsfysize=9cm
\hfil
%\epsffile[20 20 540 540]{slideg1p_hera_x_new.eps}
\epsffile[20 20 540 540]{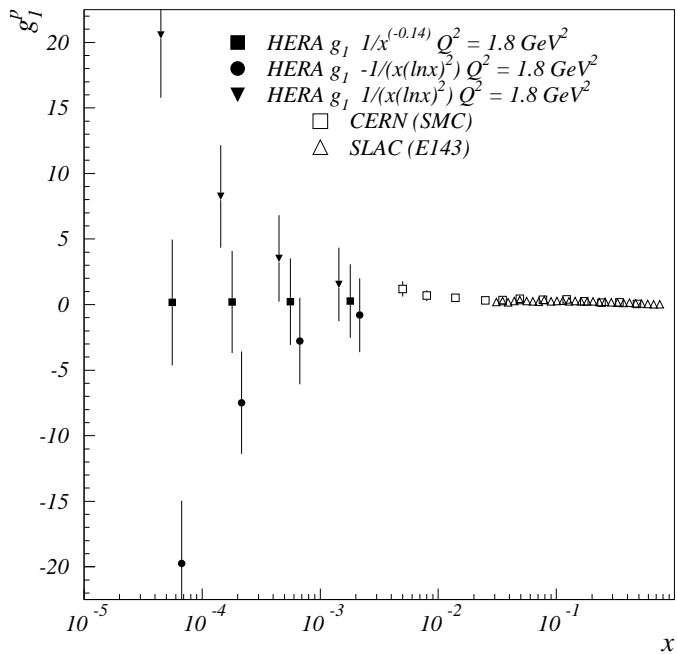}
\hfil
\caption{ {\it  The  structure function $g_{1}^{p}$ measurable at HERA
        for $Q^{2}=1.8~$GeV$^{2}$ with integrated luminosity
        $L=1000$~pb$^{-1}$ are
        shown. The SMC and E143 measurements are shown for comparison.
        Starting from the measured values of $g_{1}^{p}$ by SMC, the
        measurable values for HERA at low-$x$
        are shown in two extreme cases: valence-like behaviour
        $g_{1}(x) \sim x^{-\alpha }$ with $\alpha = 0.14$
        or strong powerlike positive or negative rises
          $g_1(x)\sim \pm 1/(x \cdot (lnx)^2)$}}
\label{xherag1p}
\end{figure}
An experimental measurement
of the small $x$ behaviour of $g_1$
would thus lead to significant insight on the structure of
QCD both within and beyond perturbation theory.
In Fig.~\ref{xherag1p} we show the expected accuracy of the determination
of $g_1$ at HERA within the maximal extent of variation in the small $x$
behaviour  compatible
with the requirement of integrability of $g_1$ (which implies
that at small $x$ $g_1$ can
rise at most as $1/(x \ln^\alpha x)$ with $\alpha>1$).
%Even though for
%definiteness only the case $g_1>0$ at small $x$ is displayed in the figure, as
%discussed above perturbative
%QCD leads to expect a rise within this range but with $g_1<0$.

\begin{figure}
\epsfxsize=10.0cm
\epsfysize=10.0cm
\hfil
\epsffile{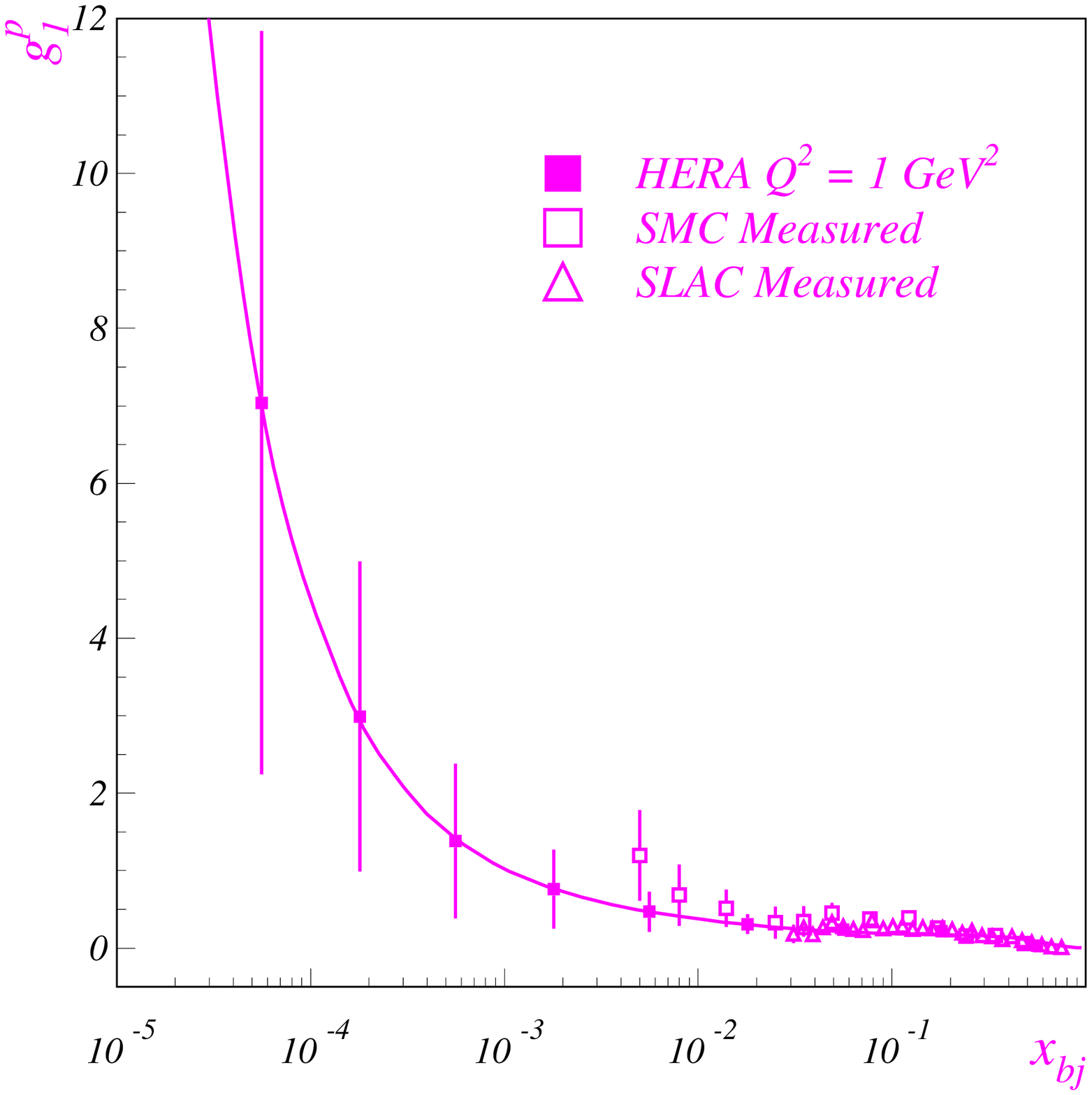}
%\epsffile{g1p_hera_smc_slac_jech.eps}
\hfil
\caption{ {\it  The starting  parametrization of $g_{1}^{p}$ at
$Q^{2}=1$~GeV$^{2}$
which gives best-fit to present-day data is shown in the $x$ range covered
at HERA.
        Statistical errors on $g_{1}$
        after combining all measurements for each $x$ are shown for
        an integrated luminosity $L = 1000~pb^{-1}$.
        }}
\label{g1hera_q21gev}
\end{figure}
\begin{figure}
\epsfxsize=5.0in
\epsfysize=6.0in
\hfil
%\epsffile{1000g1.eps}
\epsffile{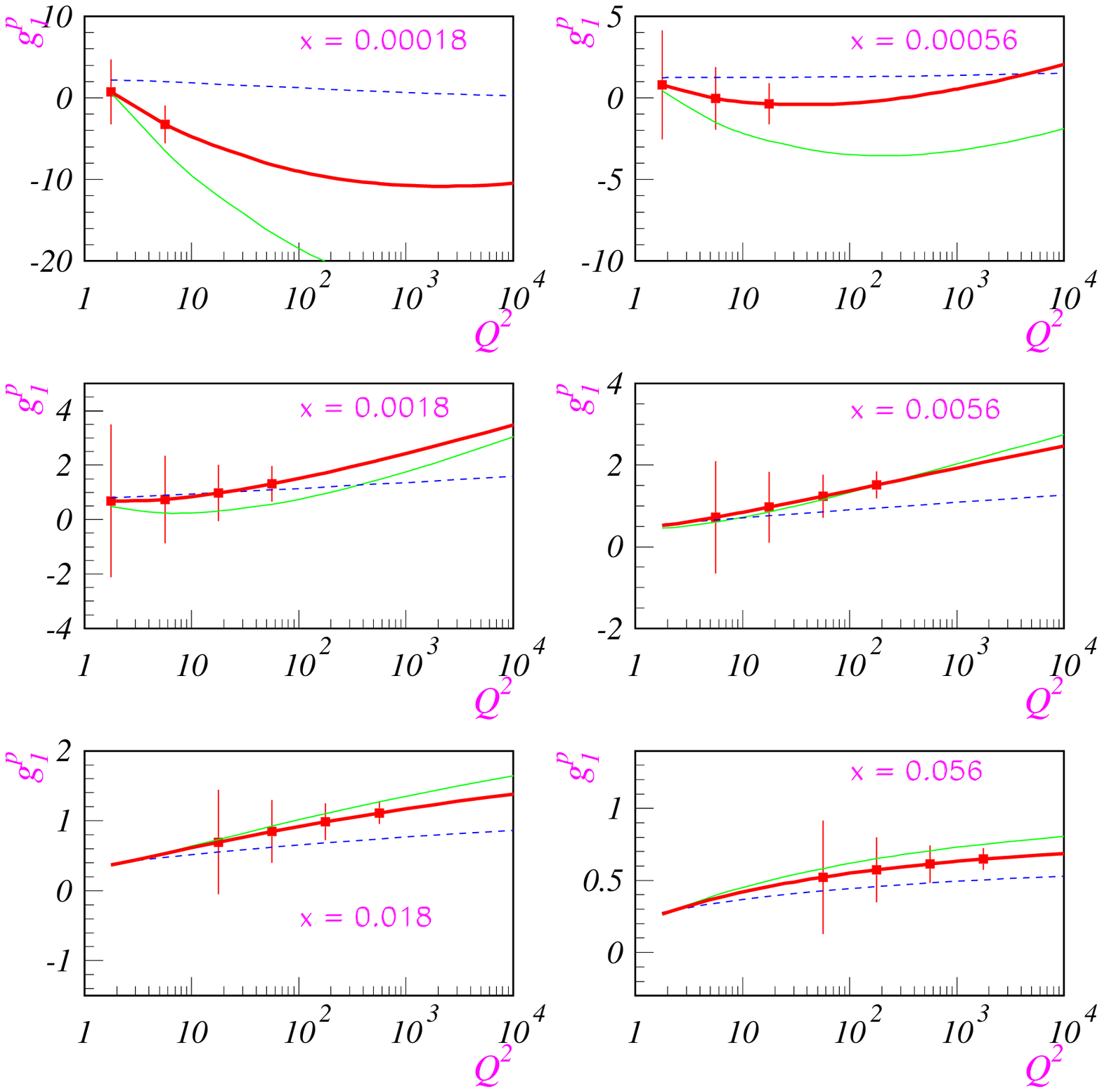}
\hfil
  \caption{ {\it Predicted values for $g^{1}_{p}$ from NLO fits together with
             estimated
             statistical uncertainties for a future polarized DIS experiment
             with an integrated luminosity L=1000~pb$^{-1}$.
             The solid bold lines are predictions based on fits to SMC and
             SLAC data extended in to the HERA kinematic range while
             the dotted and dashed lines are the maximal
             and minimal gluon predictions (see text).}}
\label{1000g1}
\end{figure}
\begin{figure}
%\caption {The NLO fit to
%deuteron $g_1^{\rm d}$ data. The solid lines are fits to data at the
%measured $Q^{2}$ values, and
%the dashed and the dotted lines are fits evolved to $Q^{2}=$ 1 and 10 GeV$^{
%2}$ repectively.}
%\epsfxsize=6.0in
%\epsfysize=6.0in
\epsfxsize=10cm
\epsfysize=10cm
\hfil
\epsffile{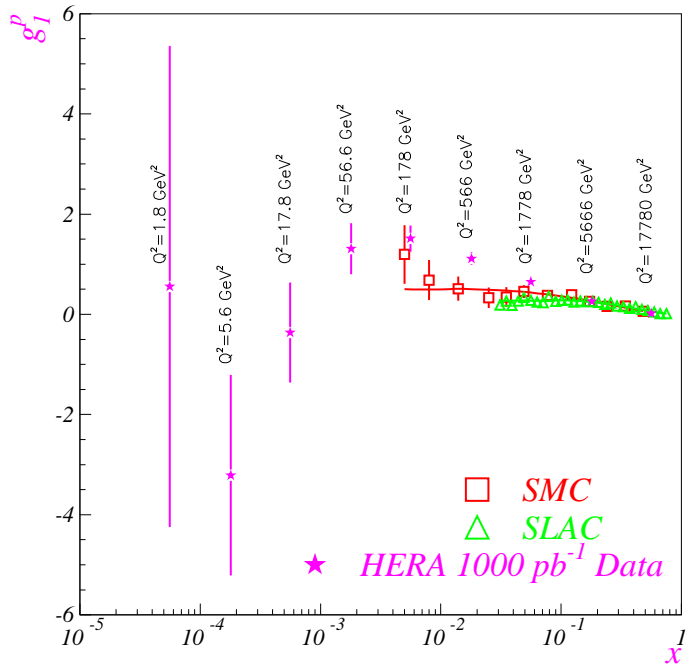}
%\epsffile{g1hera_avq2_measq2.eps}
%\epsffile{g1p_hera_smc_slac_jech.eps}
\hfil
\caption{ {\it  The structure function $g_{1}^{p}$  measurable
        at HERA shown in comparison to the SMC/E143 measurements.
        The values of $g_{1}^{p}(x,Q^{2})$ for each measurable HERA point
               are taken from
        the NLO fit and evolved to the  $Q^{2}$ value indicated in the figure.
        Statistical errors on $g_{1}(x)$ averaged over all $Q^{2}$
        for measurements at
        HERA with integrated luminosity L = 1000~pb$^{-1}$ are shown for each
        $x$.}}
\label{g1hera_q2meas}
\end{figure}
\begin{figure}
\epsfxsize=10cm
\epsfysize=10cm
\hfil
\epsffile{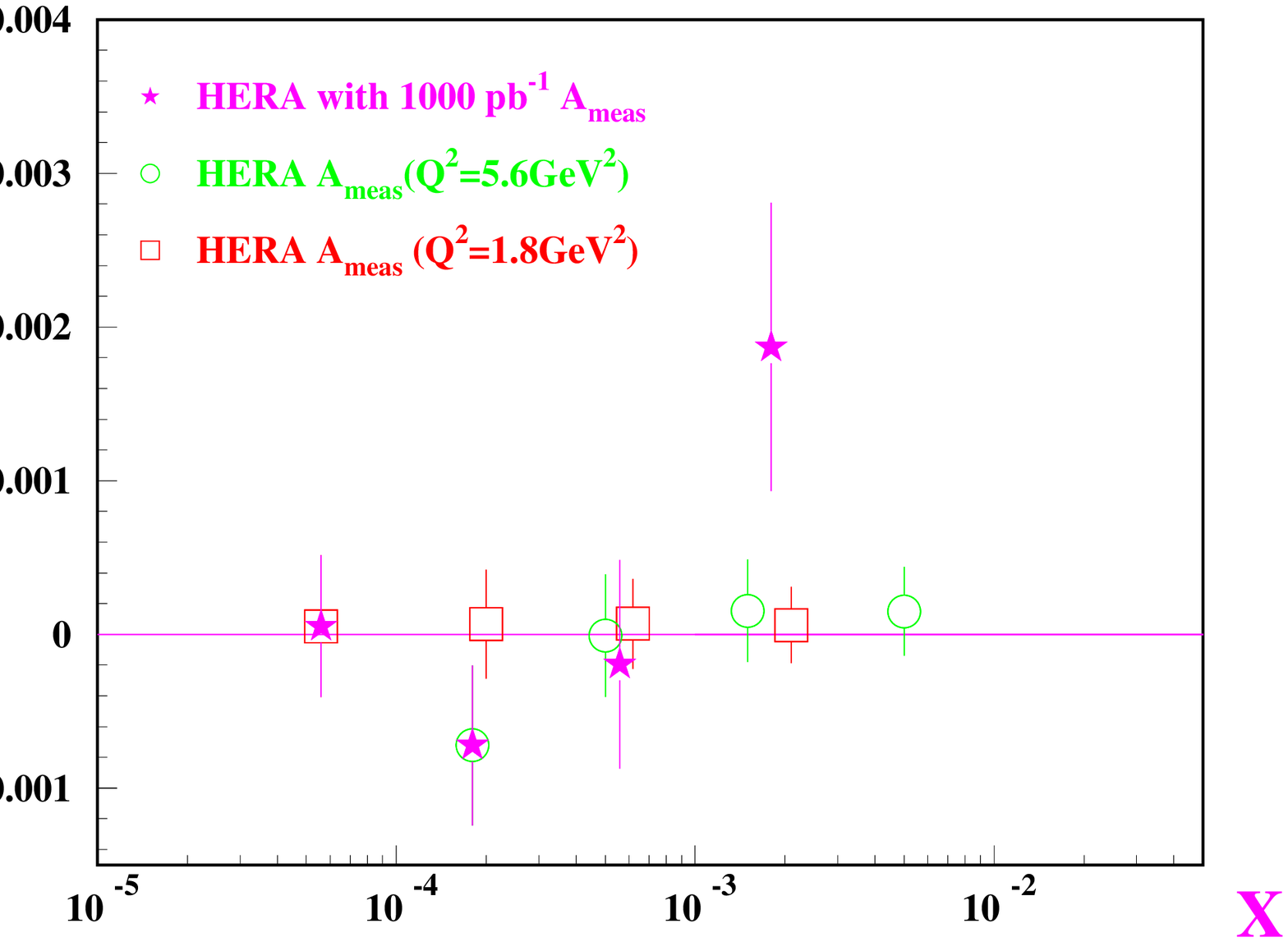}
\hfil
\caption{ {\it  The measured asymmetry $A_{m}$ for different $Q^{2}$ at HERA
for the
        projected data points shown in previous figures.} }
\label{am_hera}
\end{figure}

The full coverage of HERA experiments is shown in
Figs.~\ref{g1hera_q21gev}, \ref{1000g1}, \ref{g1hera_q2meas},
and \ref{am_hera},
where projected data with their estimated errors are given on the basis of the
NLO fit of Table~2.
The errors are estimated assuming integrated luminosities
$L=1000~pb^{-1}$
and
beam polarizations $P_p = P_e = 0.7$.
%In Fig. \ref{g1hera_q21gev} the best-fit starting parametrization of
%$g_{1}$ at 1~GeV is shown.
In Fig.~\ref{g1hera_q21gev} $g_{1}$ at the starting scale $Q^2=1$~GeV$^2$
is shown using the best fit
values of the parameters.
This is then evolved up with different
choices for the normalization of the polarized gluon distribution, to
give a feeling for the possible range of variation. In particular,
we consider two cases: a)  the first moment
of the gluon distributions is fixed to be 0  at
Q$^{2}$ = 1 GeV$^2$ (minimal gluon: dashed lines in Fig.\ref{1000g1}) and  b)
the first
moment of the singlet quark density was fixed to $\eta_q = a_{8}$ at
the same reference scale (maximal gluon: dotted lines in Fig.\ref{1000g1}).
Even though the current best fit value of $\eta_g$ (Table~2) is very close
to the maximal case, the minimal gluon is at present only excluded at
$2~\sigma$ level. Of course yet wider deviations from these fits are forseeable
since the small $x$ behaviour of the current best fit is only very loosely
constrained due to the lack of direct experimental information
at small $x$; also, as discussed above, higher order corrections beyond NLO
may turn out to be important at very small $x$.

%\begin{figure}
%\epsfxsize=10cm
%\epsfysize=10cm
%\hfil
%\epsffile{g1p_hera_highx.eps}
%\hfil
%\caption{ {\it  \label{g1phera_lowx} The same as above except that we show the
%       details of the high $x$ region.}}
%\end{figure}
%\begin{figure}
%\epsfxsize=10cm
%\epsfysize=10cm
%\hfil
%\epsffile{a1_hera.eps}
%\hfil
%\caption{ {\it  \label{a1p_hera} The  asymmetry $A_{1}^{p}$ measured at
%%different
%        $Q^{2}$ at HERA shown in comparison to the SMC/E143 measurements.
%        The values of $A_{1}^{p}$ for each HERA point was taken from
%        the NLO fits to the $g_{1}^{p}$ measured by SMC/SLAC.}
%\end{figure}
%\begin{figure}
%\epsfxsize=10cm
%\epsfysize=10cm
%\hfil
%\epsffile{a1_hera_lowx.eps}
%\hfil
%\caption{ {\it  The same $A_{1}^{p}$ as shown in the previous figure, except
%%now we
%       show the low-$x$ region.}
%\label{a1p_hera_lowx}
%\end{figure}
In Fig. \ref{1000g1} we show
the statistical errors for each measurable data point in the complete $x-Q^{2}$
grid
indicated in Fig. \ref{xq2_hera} and in Table 1. The bold solid lines are
the predictions of $g_{1}$ values in the HERA kinematic range using the
best fit values of the NLO parameters for the presently
published data~\cite{SMC_papers,E143_papers}.
%The minimal and the maximal gluon
%predictions are indicated by the dashed and dotted lines, respectively.
In Fig. \ref{g1hera_q2meas}
the projected values of $g_{1}$ obtained from the NLO fit to the data
are calculated for the lowest $Q^{2}$ data point reached at that $x$ bin, which
in turn has the lowest statistical error. The $Q^{2}$ values are indicated
in the figure.
The measured asymmetries $A_{m}$ and the
corresponding statistical errors on the projected measurements at HERA
at different $Q^{2}$ are shown in Fig.~\ref{am_hera}.

\subsection{The determination of the gluon distribution}

The measurement
of scaling violations in inclusive structure functions provides a
theoretically clean determination of the
polarized gluon distribution. Because the gluon distribution is only
determined by the scale dependence of the moments of $g_1$ a reasonably
wide kinematic coverage is required in order to achieve such a determination
with satisfactory accuracy: in particular, since the gluon distribution
is peaked at small $x$, data in this region for several values of $Q^2$
such as those obtainable at HERA would substantially improve the determination
of $\Delta g(x,Q^2)$.

To assess the impact of these data
we have repeated the fit described in sect.~3.2 with the addition of
the projected HERA data discussed in sect.~2.1.
The values of the best-fit parameters of course
do not  change, but using the
estimated errors from acceptance considerations on measurable $g_{1}$,
we get an estimate
of the extent of reduction in the measured uncertainties
of various parameters. In column 3 of Table \ref{tab-results} we
show the results of a fits with HERA data for integrated luminosity
$L=1000~pb^{-1}$ and in column 4 we show the results for $L=200~pb^{-1}$.
Note in particular the sizable improvement in the determination
of the first
moment of $\Delta g$, which is of greatest theoretical interest due
to its role in the understanding of the proton spin structure~\cite{spinrev}:
from
$\delta(\Delta g^1) = \pm 0.56$ (the present value) to $\pm 0.22$ or
$\pm 0.28$ depending
on the luminosity  available at HERA.
\begin{table}
\hfil
\begin{tabular}{|c|c|c|c|}
\hline\hline
Parameter & Published data & HERA $L=1000~pb^{-1}$ & HERA $L=200~pb^{-1}$ \\
\hline\hline
$\eta_g $ & 1.30$\pm$0.56 & 1.29$\pm$0.22 & 1.29$\pm$0.28 \\
$\eta_q $ & 0.45$\pm$0.05 & 0.45$\pm$0.04 & 0.45$\pm$0.05 \\
$\eta_{NS}$ & Fixed & Fixed & Fixed \\
\hline
$\alpha_{g}$  & -0.64$\pm$0.26 & -0.64$\pm$0.11& -0.64$\pm$0.16 \\
$\alpha_{q}$  &  0.42$\pm$0.35 & 0.42$\pm$0.26 & 0.42$\pm$0.30 \\
$\alpha_{NS}$ & -0.73$\pm$0.14 & -0.73$\pm$0.10& -0.73$\pm$0.12 \\
\hline
$\beta_{g}$   &  4.0 (fixed)  & 4.0 (fixed) & 4.0 (fixed) \\
$\beta_{q}$   &  3.53$\pm$0.87 & 3.50$\pm$0.81 & 3.50$\pm$1.13 \\
$\beta_{NS}$  &    2.18$\pm$0.28 & 2.10$\pm$0.27& 2.10$\pm$0.14 \\
\hline
$a_{q}=a_{g}$ &  1.2$\pm$2.8 & 1.2$\pm$2.5 & $1.2\pm$2.10 \\
$a_{NS}$ & 19.5$\pm$20.4& 19.5$\pm$14.0 & 19.5$\pm$16.4 \\
\hline\hline
\end{tabular}
\hfil
\caption{ {\it   Results of NLO fits: Column 2: fit to all
available published data [6, 7, 29];
 Column 3: estimated results including
data at HERA with integrated
luminosity $L=1000~pb^{-1}$;  Column 4: estimated results using
 data at HERA with
$L=200~pb^{-1}$. The parameter $\eta_{NS}$ is fixed by the octet
hyperon $\beta$ decay constant.}}
\label{tab-results}
\end{table}

%\newpage
\section{Conclusions}
Inclusive DIS measurements at HERA with high integrated luminosity
%of polarized inclusive DIS
with high energy polarized electrons and polarized protons would yield
significant and unique new information on $g_{1}^{p}(x,Q^{2})$ over a much
extended range in $x$ and $Q^{2}$. For such measurements the false asymmetries
should be considerably smaller than the true asymmetries to be measured, and
normalization systematic errors should be controlled to be less than $10\%$.
Statistical errors on the points will dominate.

There is a strong and broad current interest in the spin structure of the
nucleon. Proposals have been made and experiments are planned to
study this problem at several accelerator facilities. These include COMPASS at
CERN, a possible experiment at SLAC, and RHIC SPIN at BNL. A principal goal
of all of these proposed experiments is to measure the polarized gluon
content in the nucleon from a study of semi-inclusive processes. Another
experiment
involving a semi-inclusive process is discussed in this volume where the
determination of $\Delta g$ from a study of dijets from polarized e-p
collisions
at HERA is described. A measurement of $\Delta g$ from inclusive
process has the advantage of being theoretically clean, but
requires a wide kinematic coverage in $x$ and $Q^2$, and could thus only
be performed at HERA.

The inclusive measurements with HERA discussed in this chapter will
also provide unique information on the behaviour  $g_{1}^{p}(x,Q^{2})$
over an unexplored kinematic range, which is of great theoretical interest.

The fact that HERA is an operating facility with a plan for substantial
increase in luminosity and with two major operating detectors, as well
as a polarized electron beam, means that only the high energy polarized
proton beam needs to be developed. The efforts to achieve that seem well
justified and data from
such experiments at HERA would provide unique and complementary information
to that from other presently proposed experiments.

%\newpage

\section*{Appendix: Kinematics of polarized DIS at HERA}
\label{Appendix}
\begin{figure}
\epsfxsize=3.5in
\epsfysize=3.5in
\hfil
%\epsffile[0 0 530 530]{Dyplot.eps}
\epsffile[0 0 530 530]{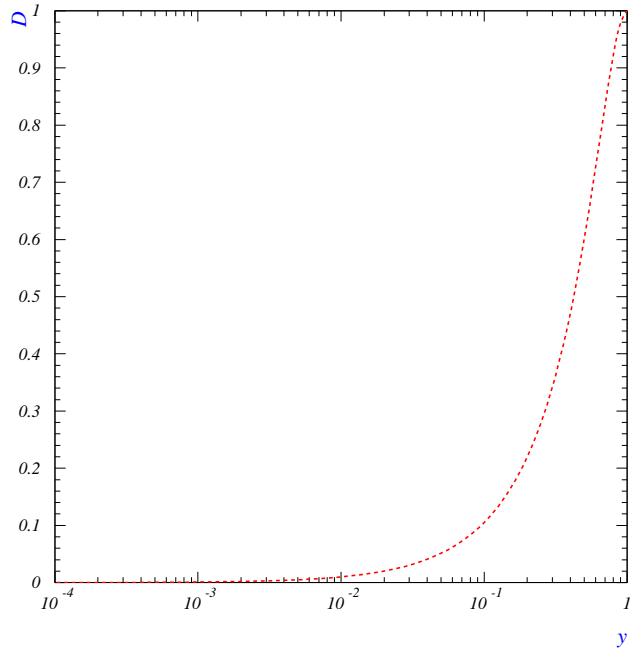}
\hfil
\caption{ {\it   The dependence of depolarization factor $D$ on $y$.
}}
\label{Dyplot}
\end{figure}

The polarized structure function $g_{1}^{p}$ of the proton is related to
the virtual photon asymmetry $A_{1}^{p}$  by the relation~\cite{yaledis},
\begin{equation}
\label{eq-g1}
g_{1} = \frac{F_{2}}{2x(1+R)} A_{1}
\end{equation}
in terms of the unpolarized structure function
$F_{2}$ and the ratio of longitudinal to transverse photoabsorption cross
sections $R$.\footnote{This result holds at leading twist.}.
The measured longitudinal asymmetry $A_{m}$ is related to
$A_{1}$ by,
\begin{equation}
\label{eq-A}
A_{m} = \frac{N^{\uparrow \uparrow} - N^{\uparrow \downarrow}}
{N^{\uparrow \uparrow} + N^{\uparrow \downarrow}} = P_{p}P_{e}DA_{1}
\end{equation}
where $P_{p}(P_{e})$ is the proton (electron) polarization, and D is the
depolarization factor.
The arrows
indicate the
relative direction of the spins of the electrons and protons.
$D$ is calculable from QED and is given by
\begin{equation}
\label{eq-Depol}
D = \frac{y(2-y)}{y^{2}+2(1-y)(1+R),}
\end{equation}
where
\begin{equation}
y=\frac{Q^{2}}{ x s}
\end{equation}
and $s$ is the Mandelstam invariant of the electron-proton collision.
The dependence of $D$ on $y$ is shown in
Fig.~\ref{Dyplot}.
%\begin{figure}
%\epsfxsize=3.5in
%\epsfysize=3.5in
%\hfil
%\epsffile{x_q2_y_plot.eps}
%\hfil
%\caption{ {\it  The $x-Q^{2}$ plane and the lines of constant $y$ at HERA
%%kinematics.}
%\end{figure}
The statistical uncertainty $\delta A_{m}$ in the measurement of $A_{m}$
is given by,
\begin{equation}
\label{eq-dA}
\delta A_{m} = \frac{1}{\sqrt{N^{\uparrow \uparrow}+N^{\downarrow \uparrow}}}
= \frac{1}{\sqrt{N_{total}}}
\end{equation}
where $N^{\uparrow \uparrow~(\uparrow \downarrow)}$ represent the number of
DIS events observed with parallel
(anti parallel) proton and electron polarizations, respectively.
The sum in the
denominator is then the square root of the total number of events observed
in the experiment
at selected $x-Q^2$ bins.

We estimate the yield and the associated statistical error to be observed
in a future HERA experiment.
The cross section for DIS is given by
\begin{equation}
\label{eq-dis_cs}
\frac{d^{2}\sigma}{dxdQ^{2}}=\frac{4\pi \alpha^{2}}{xQ^{4}}
\left[1-y+\frac{y^2}{2(1+R)}\right]F_{2}
\end{equation}
The depolarization factor $D$ for each of the bins was calculated and
used to estimate the statistical uncertainty in the measurement of
$g_{1}^{p}$ from Eqs. \ref{eq-Depol} and \ref{eq-dA}.

{\bf Acknowledgements:}
We thank G. Altarelli and P. Schuler for interesting discussions and helpful
comments.

\end{document}